# Chlorine and Bromine Isotope Fractionation of Halogenated Organic Compounds in Electron Ionization Mass Spectrometry

**Caiming Tang[1,2,*], Jianhua Tan[3], Zhiqiang Shi[4], Caixing Tang[5], Songsong Xiong[1,2], Jun Liu[1,2], Yujuan Fan[1,2], Xianzhi Peng[1,*]**

[1] *State Key Laboratory of Organic Geochemistry, Guangzhou Institute of Geochemistry, Chinese Academy of Sciences, Guangzhou 510640, China*

[2] *University of Chinese Academy of Sciences, Beijing 100049, China*

[3] *Guangzhou Quality Supervision and Testing Institute, Guangzhou, 510110, China*

[4] *College of Chemistry and Chemical Engineering, Taishan University, Taian 271021, P. R. China.*

[5] *The Third Affiliated Hospital of Sun Yat-sen University-Lingnan Hospital, Guangzhou 510630, China*

\*Corresponding Author.

Tel: +86-020-85291489; E-mail: CaimingTang@gig.ac.cn. (C. Tang).

Tel: +86-020-85290009; Fax: +86-020-85290009. E-mail: pengx@gig.ac.cn. (X. Peng).

# ABSTRACT

Revelation of chlorine and bromine isotope fractionation of halogenated organic compounds (HOCs) in electron ionization mass spectrometry (EI-MS) is crucial for compound-specific chlorine/bromine isotope analysis (CSIA-Cl/Br) using gas chromatography EI-MS (GC-EI-MS). This study systematically investigated chlorine/bromine isotope fractionation in EI-MS of HOCs including 12 organochlorines and 5 organobromines using GC-double focus magnetic-sector high resolution MS (GC-DFS-HRMS). Chlorine/bromine isotope fractionation behaviors of the HOCs in EI-MS showed varied isotope fractionation patterns and extents depending on compounds. Besides, isotope fractionation patterns and extents varied at different EI energies, demonstrating potential impacts of EI energy on the chlorine/bromine isotope fractionation. Hypotheses of inter-ion and intra-ion isotope fractionations were applied to interpreting the isotope fractionation behaviors. The inter-ion and intra-ion isotope fractionations counteractively contributed to the apparent isotope ratio for a certain dehalogenated product ion. The isotope fractionation mechanisms were tentatively elucidated on basis of the quasi-equilibrium theory. In the light of the findings of this study, isotope ratio evaluation scheme using complete molecular ions and the EI source with sufficient stable EI energies may be helpful to achieve optimal precision and accuracy of CSIA-Cl/Br data. The method and results of this study can help to predict isotope fractionation of HOCs during dehalogenation processes and further to reveal the dehalogenation pathways.

# INTRODUCTION

Gas chromatography coupled with electron ionization quadrupole mass spectrometry (GC-EI-qMS) has been recently employed to conduct compound-specific chlorine/bromine isotope analysis (CSIA-Cl/Br) of halogenated organic compounds (HOCs) such as tetrachloroethene, trichloroethene, cis-dichloroethene,[1] p,p'-dichlorodiphenyltrichloroethane (p,p'-DDT), pentachlorophenol,[2] chlorinated acetic acids,[3] bromoform, 3-bromophenol and 4-bromotoluene.[4] Performances of CSIA-Cl methods using GC-EI-qMS were found to be comparable with those of conventional methods using GC continuous flow isotope ratio MS (GC-IRMS) and preparative GC followed by offline IRMS.[3,5] Moreover, CSIA-Cl methods using GC-qMS significantly simplify sample pretreatment and therefore are more cost-effective and applicable.[2,5] As a result, application of GC-EI-MS in CSIA-Cl studies has been attracting increasing interests hitherto.[6-9] A recent study reported CSIA-Cl of hexachlorobenzene (HCB) in air samples using GC quadrupole hybrid time-of-flight (TOF) high resolution MS (GC-QTOF-HRMS) equipped with an EI source[10] with an similar isotope-ratio evaluation scheme as those applied in the studies using GC-qMS.[1,5,11]

However, some intrinsic limitations of GC-EI-MS may trigger uncertainties and deviations in CSIA results and therefore should be ascertained and avoided. In a previous study, we found that HOCs could exhibit chlorine/bromine isotope fractionation on GC columns, probably leading to biased CSIA-Cl/Br results.[12] In addition, chemical bonds may be cleaved (ion fragmentation in ion source and metastable-ion disassociation in EI-MS) or formed in EI-MS, which means that kinetic isotope effects (KIEs) may occur, subsequently leading to observable isotope fractionation and resulting in imprecision and inaccuracy of CSIA-Cl results.[10] As a result, suitable external isotopic standards which are chemically identical to target analytes are generally needed for CSIA-Cl to correct deviations caused by possible isotope fractionation in EI-MS.[2] In a review about hydrogen/deuterium isotope effects during fragmentation in EI-MS, both intramolecular and intermolecular isotope effects could take place in EI-MS.[13] An approximation scheme method for calculating KIE was also proposed based on mass spectrometric signal intensities of the dehydrogenation and de-deuterium product ions. The hydrogen/deuterium isotope effects were "staggeringly large"[13] with KIEs reaching up to two or three orders of magnitude ($10^2$-$10^3$).[14,15] In addition, significant intramolecular hydrogen/deuterium isotope effects of aniline and protonated aniline were observed in high pressure chemical ionization MS and the metastable ion fragmentation and the adduct ion collision-induced dissociation processes also presented high intramolecular hydrogen/deuterium isotope effects.[16] Therefore, both inter-ion and intra-ion Cl/Br isotope fractionations may be anticipated to occur for HOCs during fragmentation processes in EI-MS, which could lead to complex and hard-to-predict isotope fractionation modes. However, to the best of our knowledge, Cl/Br isotope fractionation in EI-MS and the potential impacts on CSIA results have not been revealed in detail yet.

Herein, we used GC-EI-double focus magnetic sector HRMS (GC-EI-DFS-HRMS) to investigate Cl/Br isotope fractionation of 17 HOCs, including 12 organochlorines and 5 organobromines during dehalogenation process in EI-MS. Hypotheses of inter-ion and intra-ion isotope fractionations were proposed to interpret the observed isotope fractionation patterns. Mechanisms of the isotope fractionation were also tentatively elucidated on basis of the quasi-equilibrium theory (QET).[17]

## EXPERIMENTAL SECTION

Information about the *chemicals* (names and abbreviations) and *materials* are provided in the *Supporting Information*.

**GC-HRMS Analysis.** The GC-HRMS consisted of dual Trace-GC-Ultra GC equipped with a DFS-HRMS and a TriPlus auto-sampler (GC-DFS-HRMS, Thermo-Fisher Scientific, Bremen, Germany). A DB-5 MS capillary column (60 m × 0.25 mm, 0.25 μm thickness, J&W Scientific, USA) was employed for separation. The GC oven temperature programs are provided in Table S-1. The working solutions were analyzed by GC-HRMS directly.

Positive EI source was applied with energies set at 30, 45, 55 or 70 eV. Ion source temperature was 250 °C. Data acquisition was carried out in multiple ion detection (MID) mode. A dwell time of around 15 ms was set for each isotopologue ion. The mass resolution (5% peak-valley definition) was ≥ 10000 and the MS detection accuracy was set at ±0.001 u. MS was calibrated with perfluorotributylamine in real time during MID acquisition.

Chemical structures of the HOCs and their dehalogenated radical fragments were depicted with ChemDraw (Ultra 7.0, Cambridgesoft). The exact masses of the Cl/Br isotopologues of parent molecules and their dehalogenated radical fragments were thus calculated with accuracy of 0.00001 u. For a compound or a dehalogenated radical fragment containing $n$ Cl/Br atoms, $n+1$ isotopologues were selected. The mass-to-charge ($m/z$) values of isotopologues of the molecular ions and dehalogenated product ions can be calculated by subtracting the mass of an electron from the calculated exact masses. The $m/z$ values were then imported into MID modules to detect the molecular ions and the dehalogenated product ions. Information about isotopologues of the investigated compounds along with their dehalogenated radical fragments, including chemical formulas, isotopologue formulas, and exact *m/z* values are provided in Table S-2.

**Data Processing.** Isotope ratio (IR) of the isotopologue ions of each compound or dehalogenated fragment was calculated as:

$$IR = \frac{\sum_{i=0}^{n} i \cdot I_i}{\sum_{i=0}^{n} (n-i) \cdot I_i} \quad (1)$$

where $n$ is number of Cl/Br atom(s) of a molecular ion or a dehalogenated product ion; $i$ is number of $^{37}$Cl or $^{81}$Br atom(s) in an isotopologue ion; $I_i$ is the MS signal intensity of the isotopologue ion $i$. This calculation method is defined as complete-isotopologue scheme of isotope ratio evaluation.

The isotope ratio of the total ions (all the measured molecular and product ions were involved, IR$_{total}$) of each compound was calculated as:

$$IR_{total} = \frac{\sum_{n=1}^{m} \sum_{i=0}^{n} i \cdot I_i}{\sum_{n=1}^{m} \sum_{i=0}^{n} (n-i) \cdot I_i} \quad (2)$$

where $m$ is the number of Cl/Br atoms in the molecular formula of a compound. This method is defined as total-ion scheme of isotope ratio evaluation.

All the measured isotope ratios were relative values instead of the true isotope ratios referenced with the SMOC/B (Standard Mean Ocean Chlorine/Bromine) values, due to unavailability of external isotopic standards of the investigated HOCs,.

Ionization and fragmentation processes of HOCs in EI-MS were hypothesized to be in the following sequence: ionization of molecule → molecular ion → dehalogenation by losing Cl/Br atom(s) one by one → production of a series ions with successive numbers (from 1 to *m*) of Cl/Br atom(s). An ion with *n* Cl/Br atoms of a compound was defined as the precursor ion of the ion with *n*-1 Cl/Br atom(s), and the carbon skeleton remained intact, and the latter was defined as the product ion of the former. For example, the ions [M]$^+$ and [M–Cl]$^+$ were regarded as the precursor ions of [M–Cl]$^+$ and [M–2Cl]$^+$, and the ions [M–Cl]$^+$ and [M–2Cl]$^+$ were defined as the product ions of [M]$^+$ and [M–Cl]$^+$, respectively.

According to the QET,[17] if the internal energy and time scale in mass spectrometry are fixed for a compound, the fractions of all the ions derived from this compound are predictable certain values, which means the relative abundances of all the ions can reach an equilibrium state. Therefore, in this study, we deemed the relative abundances of isotopologue ions of a compound reached an equilibrium state. Accordingly, isotope fractionation extent from a precursor ion to its product ion was evaluated with equilibrium fractionation factor ($\alpha_{eq}$) and enrichment factor ($\varepsilon$)[18]:

$$\alpha_{eq} = \frac{IR_{Product}}{IR_{Precusor}} \quad (3)$$

where $IR_{Product}$ and $IR_{Precusor}$ are isotope ratios of the product ion and the corresponding precursor ion, respectively.[19] For simplifying evaluation of the observed isotope fractionation, further isotope fractionation was not taken into account during subsequent dehalogenation of the product ion.

When $\varepsilon < 0$, the isotope ratio of the precursor ion is higher than its product ion, indicating enrichment of heavy isotope in the precursor ion and deficit in the product ion. On the other hand, when $\varepsilon > 0$, the isotope ratio of the precursor ion is lower than its product ion, demonstrating deficit of heavy isotope in the precursor ion and enrichment in the product ion. The measured values of $\alpha_{eq}$ and $\varepsilon$ show the apparent isotope fractionation from precursor ions to the corresponding product ions.

Data processing procedures were illustrated in Figure 1 taking o,p'-DDE as an example. First, the mass spectrum of o,p'-DDE was extracted from the total ion chromatogram, giving rise to the average MS signal intensity of each ion within the retention time range.

Then, the isotope ratios of the molecular ion, the dehalogenated product ions, and the total ions were calculated with the obtained average intensities of isotopologue ions by eq 1 or eq 2. Data derived from five replicated injections were employed to calculate the mean isotope ratios and standard deviations (1σ).

## RESULTS AND DISCUSSION

**Evaluation Schemes for Chlorine/Bromine Isotope Fractionation.** Complete-isotopologue and total-ion schemes were applied to evaluation of isotope ratios. The former was based on MS signal intensities of the isotopologue ions (ions with the same number of Cl/Br atom(s)) as indicated in eq 1, whereas the latter calculated isotope ratios with signal intensities of all monitored ions as shown in eq 2. The measured isotope ratios along with the corresponding standard derivations of the molecular ions and the dehalogenated product ions were employed to evaluate the isotope fractionation. All the isotope ratios measured in this study were relative values of isotopologue ions due to unavailability of external isotopic standards. However, all the isotopolugue ions were detected simultaneously, leading to accurate relative isotope ratios for the molecular ion and dehalogenated product ions. Therefore, the quotient of the relative isotope ratio of a product ion to that of its precursor ion, i.e., the $\alpha_{eq}$, was capable of reflecting the isotope fractionation from the precursor ion to the product ion in EI-MS. In addition, the ε values were calculated to more directly assess the isotope fractionation extents.

**Chlorine Isotope Fractionation.** All the 12 investigated chlorinated compounds presented isotope fractionation in EI-MS at the EI energy of 45 eV (Figure 2 and Table S-3). The isotope fractionation patterns and extents varied depending on compounds. It's worthy to note that the isotope fractionation extents were significantly different for the two DDE isomers, o,p'-DDE and p,p'-DDE(Figure 2, G2). , The chlorine isotope ratio of P-$Cl_3$ ion of o,p'-DDE was 0.4146 ±0.0084, much higher than that of its parant ion M-$Cl_4$ ion (0.3209 ±0.0009), indicating a significant chlorine isotope fractionation during dechlorination from M-$Cl_4$ to P-$Cl_3$, with the ε as high as 292.0‰. On the other hand, the chlorine isotope ratio of P-$Cl_3$ ion of p,p'-DDE was 0.3288 ± 0.0049 (ε = 25.0‰), significantly lower than that of o,p'-DDE, but still evidently higher than that of its precursor ion M-$Cl_4$ (0.3208 ±0.0012). The chlorine isotope ratios of M-$Cl_4$, P-

$Cl_2$, and P-$Cl_1$ of the two DDE isomers were pairwise similar. The difference in the extents of isotope fractionation was probably attributable to the large isotope ratio discrepancy between the two P-$Cl_3$ ions generated by losing a Cl atom from the corresponding molecular ions (M-$Cl_4$).

On the contrary, PCB-18, PCB-28, o,p'-DDT and p,p'-DDT showed similar inter-isomer patterns and extents of isotope fractionation (Figure 2, G1 and G3). For instance, the chlorine isotope ratios of PCB-18 and PCB-28 declined with decreasing Cl atoms on the ions, demonstrating heavy-isotope enrichment in the precursor ions. While DDTs generally exhibited decreasing isotope ratios from the precursor ions to the product ions (Figure 2, G3) with the exception of P-$Cl_3$→P-$Cl_2$ ($\varepsilon$ of -20.55 and -21.54 ). For the other organochlorine compounds, however, no clear relationship was observed between the isotope fractionation extents and the number of Cl atoms on the ions. Most of the investigated organochlorine compounds, except the Tri-PCBs and DDTs, presented fluctuated isotope fractionation patterns versus the number of Cl atoms of the ions, typically for HCB (Figure 2, G5). The isotope fractionation directions of these compounds changed alternately as indicated by the values of $\alpha_{eq}$ and $\varepsilon$ of these compounds in Table S-3.

The absolute values of most $\varepsilon$ values were larger than 20‰, indicating significant chlorine isotope fractionation during dechlorination reactions of the investigated chlorinated compounds in EI-MS compared with the reported values during biotic and abiotic transformations of chloroethylenes (generally less than 5‰),[18] especially o,p'-DDE and HCB (Table S-3).

The two DDTs and o,p'-DDD have a chloromethyl (dichloromethyl or trichloromethyl) in the structures which can result in intensive dechloromethylation product ions in EI-MS. Differences in the chlorine isotope ratios were investigated between the dechlorinated product ions and dechloromethylated product ions of these compounds. The chlorine isotope ratios of P-$Cl_2$–CM ions (dechloromethylated product ions with two Cl atoms) of the three compounds were evidently higher than those of the P-Cl$_2$ ions (Figure 3). The P-Cl$_2$–CM ions had two Cl atoms on the benzene rings, while the P-Cl$_2$ ions might possess Cl atoms on the benzene rings and/or on the chloromethyl. According to the reported bond dissociation energies, strength of a C-Cl bond is higher on a benzene ring than on a methyl,[20] which means that the Cl atoms on a benzene ring are less likely to be cleaved than those on a chloromethyl. It can therefore be speculated that the Cl atoms of P-Cl$_2$ ions were mainly on the benzene rings. The chlorine isotope ratios of the chloromethyls might be lower than those of the chlorinated benzene rings, which leaded to attenuation of $^{37}$Cl content of the P-Cl$_2$ ions. Chlorine isotope ratios ascended from P-Cl$_2$ ions to P-Cl$_1$ ions with enrichment factors of 35.6‰, 68.6‰ and 59.7‰ for o,p'-DDD, o,p'-DDT and p,p'-DDT, respectively (Table S-3). The chlorine isotope ratios of the P-Cl$_1$ ions and the P-Cl$_2$–CM ions of the two DDTs were similar (Figure 3). Besides, no significant isotope fractionation was found from P-Cl$_2$–CM ions to P-Cl$_1$–CM ions for two DDTs. For o,p'-DDD, however, the chlorine isotope ratio of P-Cl$_1$ ion was evidently higher than that of P-Cl$_1$–CM ion. Furthermore, evidently decline of chlorine isotope ratios was observed from P-Cl$_2$–CM ion to P-Cl$_1$–CM ion. Theoretically, P-Cl$_2$–CM and P-Cl$_1$–CM ions of o,p'-DDD are structurally identical to those of o,p'-DDT. Mechanisms generating different isotope fractionation modes from the P-Cl$_2$–CM ions to the P-Cl$_1$–CM ions between o,p'-DDD and o,p'-DDT were not clear and more research is needed in the future.

**Bromine Isotope Fractionation.** All the five investigated organobromine compounds (i.e., TBDD, TBDF, BDE-77, BDE-118, and $^{13}$C$_6$-HBB) exhibited obvious bromine isotope fractionation in EI-MS at 45 eV (Figure 2, G6-8). TBDD and BDE-77 showed similar isotope fractionation patterns with pairwise comparable bromine isotope ratios of the M-Br$_4$, P-Br$_2$ and P-Br$_1$ ions. However, bromine isotope fractionation patterns between TBDD and TBDF were different despite similarity in their chemical structures. Similar Cl/Br isotope fractionation patterns were observed for HCB and $^{13}$C$_6$-HBB, with fluctuations during successive dehalogenation reactions (Figure 2, G5 and G7), probably impling identical dehalogenation pathways of the two compounds. Since HCB and $^{13}C_6$-HBB are planar molecules with highly symmetric structures, the six Cl/Br atoms could be randomly ionized with equal possibility in EI source. Once a Cl/Br atom is ionized, the electron density between the atoms changes or a chemical bond is elongated (weakened), and a Cl/Br atom will thus be more liable to be removed from the benzene ring.

**Effects of EI Energy on Isotope Fractionation.** Four different EI energies, 30 eV, 45 eV, 55 eV and 70 eV, were set to investigate the effects of EI energy on Cl/Br isotope fractionation in EI-MS with eight exemplary compounds, i.e., PCB-18, PCB-28, PCB-52, PCB-101, o,p'-DDE, p,p'-DDE, o,p'-DDD and BDE-77. As shown in Figure 4, the isotope fractionation patterns varied with EI energy for most of the compounds with exception of o,p'-DDE and BDE-77 whose isotope fractionation patterns were consistent. It's therefore speculated that the stability of EI energy may affect precision of CSIA-Cl/Br results.

**Mechanistic Exploration of Cl/Br Isotope Fractionation**

Similar to the reported intermolecular and intramolecular isotope fractionations,[13] two types of isotope fractionations, i.e., inter-ion isotope fractionation and intra-ion isotope fractionation might contribute to the apparent Cl/Br isotope fractionation observed in this study. According to the differences of zero point energies between isotopomers, a bond linking carbon atom with light isotope ($^{35}Cl$ or $^{79}Br$) is more liable to be cleaved than that linking carbon atom with heavy isotope ($^{37}Cl$ or $^{81}Br$) provided they are structurally identical. Therefore, lighter isotopologues are more likely to generate dehalogenation reactions, resulting in deficit of heavy isotope in the product ions and enrichment of heavier isotopologues in precursor ions, demonstrating the so-called inter-ion isotope fractionation. On the other hand, the heavy isotope is more liable to remain on the product ions, thus the isotopologues of the product ions are more liable to become heavier, showing intra-ion isotope fractionation. For a certain product ion generated during a dehalogenation reaction, the inter-ion and intra-ion isotope fractionations are theoretically in opposite directions, positively and negatively affecting the isotope ratio of this product ion, respectively.

*Considerations Based on the QET.* The fundamental principles inter-ion and intra-ion isotope fractionations occurring in EI-MS can be explained with the QET from energy perspective.[13] According to the QET, the rate constant $k(E)$ for a unimolecular reaction is:

$$k(E) = \frac{G^*(E - E_0)}{h\rho(E)} \qquad (4)$$

where $E$ is the internal energy of a reaction ion, $E_0$ is the critical energy of the reaction, $G^*(E–E_0)$ is the amount of internal energy states of the transition state of the ion in the energy range from $E_0$ to $E$, $h$ is the Planck's constant, and $\rho(E)$ is the state density of the reacting ion at energy $E$. If an ion lost a Cl/Br atom via two isotopically different pathways with rate constants of $k_l(E)$ and $k_h(E)$ to generate two product ions, the KIE can be expressed as:

$$KIE = \frac{k_l(E)}{k_h(E)} = \frac{G_l^*(E - E_{0l})}{G_h^*(E - E_{0h})} \frac{\rho_h(E)}{\rho_l(E)} \qquad (5)$$

where the subscripts ($l$) and ($h$) represent light and heavy isotope, respectively. Hence, the KIEs during fragmentation in EI-MS can be determined by internal energies, critical energies, the amount of internal energy states of transition state, and state density of reacting ion at energy $E$. The KIEs are excepted to >1 for both the inter-ion and intra-ion isotope fractionations.[13]

*Inter-ion Isotope Fractionation.* To elucidate the inter-ion isotope fractionations for compounds containing the isotope atoms of which all are position-equivalent (symmetric), a general equation was derived:

$$\frac{\sum_{i=0}^{n} i \cdot x_i}{\sum_{i=0}^{n} (n-i) \cdot x_i} = \frac{\sum_{i=0}^{n} \sum_{t=0}^{n-r} t \cdot C_{n-i}^{n-r-t} C_i^t \cdot x_i}{\sum_{i=0}^{n} \sum_{t=0}^{n-r} (n-r-t) \cdot C_{n-i}^{n-r-t} C_i^t \cdot x_i} \quad (6)$$

where $n$ is the number of Cl/Br atoms of a precursor ion, $i$ is the number of $^{37}$Cl or $^{81}$Br atom(s) of an isotopologue of the precursor ion, $r$ is the number of the lost Cl/Br atom(s) (for the dehalogenation process from a precursor ion to the product ion in this study, $r$ is equal to 1), $t$ is the number of $^{37}$Cl or $^{81}$Br atom(s) of an isotopologue of a product ion derived from the precursor ion, $x_i$ is the molar amount of isotopologue $i$ of the precursor ion. For a specific case, when $r$ is 1, the equation becomes:

$$\frac{\sum_{i=0}^{n} i \cdot x_i}{\sum_{i=0}^{n} (n-i) \cdot x_i} = \frac{\sum_{i=0}^{n} \sum_{t=0}^{n-1} t \cdot C_{n-i}^{n-1-t} C_i^t \cdot x_i}{\sum_{i=0}^{n} \sum_{t=0}^{n-1} (n-1-t) \cdot C_{n-i}^{n-1-t} C_i^t \cdot x_i} \quad (7)$$

The deducing and proving processes of eq 6 are provided in an unpublished manuscript.[21] As can be seen in eq 7, if intra-ion isotope effect is not taken into account, the isotope ratio calculated from MS signal intensities of the product ion isotopologues equals that of the precursor ion isotopologues. Since the lighter isotopologues are more prone to be dehalogenated in EI-MS, inter-ion isotope fractionation tends to result in enrichment of heavy isotope for the precursor ions but lead to deficit of heavy isotope for the product ions.

For an asymmetric precursor ion containing position-distinct isotope atoms, if the dehalogenation from the precursor ion to the product ion reacts at only one position, the isotope ratio of the remaining isotope atoms on the product ion equals that of the total precursor ions (both dehalogenated and non-dehalogenated) provided that the isotope ratio of the isotope atoms on the reacting position is not position-specific. The isotope ration of this product ion is anticipated to be lower than that of the remaining (non-dehalogenated) precursor ion, but higher than that of the reacting (fragmented) precursor ion. As a result, the inter-ion isotope fractionation taking place in this condition is normal for the precursor ion but without effects on the product ion.

A prerequisite of inter-ion isotope fractionation to happen is that a precursor ion is not completely dehalogenated. If all precursor ion isotopologues were dehalogenated by losing one isotope atom, the isotope ratio of the product ion was the same with that of the precursor ion (for an asymmetric compound, only if the isotope ratio of the atoms is not position-specific). In this study, all the molecular ions and dehalogenated product ions were observed, thus no reaction position lost all Cl/Br atoms during one-step dehalogenation, indicating inter-ion isotope fractionation in all dehalogenation processes in EI-MS.

The inter-ion isotope fractionation can be typically elucidated with the isotopologue ions possessing completely different isotopes, e.g., $[R^{35}Cl_2]^+$ and $[R^{37}Cl_2]^+$ (R is a hypothesized functional group). Since the KIE is > 1, during ionization and fragmentation time scale (around $10^{-5}$-$10^{-6}$ s), more $[R^{35}Cl_2]^+$ ions can be dechlorinated to $[R^{35}Cl]^+$ than $[R^{37}Cl_2]^+$ to $[R^{37}Cl]^+$. As a result, for the precursor ion, more $[R^{37}Cl_2]^+$ ions are remained than $[R^{35}Cl_2]^+$, thus the heavy isotope is enriched, demonstrating occurrence of the isotope fractionation positively affecting the isotope ratio of the precursor ion. On the other hand, the heavy isotope is diluted for the product ion ($[R^{35}Cl]^+$ and $[R^{37}Cl]^+$) as more $[R^{35}Cl]^+$ ions are produced than $[R^{37}Cl]^+$). Therefore, the inter-ion isotope fractionation affects the isotope ratios of the precursor ions positively, but impacts those of product ions negatively. In this case (isotopologues containing completely different isotopes), the precursor ion that finally transformed into the product ion has the isotope ratio equal to that of the product ion.

*Intra-ion Isotope Fractionation.* Intra-ion isotope fractionation may happen when isotopologue formula of a precursor ion contain both light and heavy isotopes and the isotopes can leave from at least two positions in generation of product ion from the precursor ion. In addition, the isotopes on either position should not be completely cleaved. The reaction positions can be structurally identical or different provided the reactions are synchronous and alternative. If the reaction positions are structurally identical, then the reactions are only isotopically different. The KIE can thus be expressed as:

$$KIE = \frac{k_l(E)}{k_h(E)} = \frac{G_l^*(E-E_{0l})}{G_h^*(E-E_{0h})} \tag{8}$$

The states density, $\rho(E)$, is the same for the two reaction pathways and thus cannot impact the intra-ion isotope fractionation.[13] This is a special case when the synchronous reaction positions are structurally identical. If the reaction positions are structurally different, critical energies of the reactions should not be significantly different, thus insuring the reactions on two identical ions can occur simultaneously and respectively.

To elucidate intra-ion isotope fractionation, we take the hypothesized ion $[R^{35}Cl^{37}Cl]^+$ for an example. Since the KIE is > 1, more $^{35}Cl$ atoms can leave the precursor ion $[R^{35}Cl^{37}Cl]^+$ than $^{37}Cl$ atoms, thus more product ions $[R^{37}Cl]^+$ are generated relative to the $[Cl_2^{35}Cl]^+$. Consequently, the ratio $[R^{37}Cl]^+ / [Cl_2^{35}Cl]^+$ is expected to be higher than that of the precursor ion, for which the isotope ratio is identically equal to 1. Thus, the intra-ion isotope fractionation leads to enrichment of heavy isotope for the product ions.

According to the QET, intra-ion KIEs significantly depend on internal energy (eq 8), presenting steady falling tendencies as internal energy increases. On the other hand, the increase of difference between the critical energies involving light isotope and heavy isotope raises intra-ion KIEs (eq 5 and 8). Theoretically, intra-ion KIEs will be infinite if an internal energy is sufficiently low to just above the critical energy of one reaction involving a light isotope and below that of another involving a heavy isotope. In addition, the probability of the internal energy coming within the energy region of the critical energies ($E_{0l}$ to $E_{0h}$) increases with the increase of the difference between the critical energies. Thus, in EI-MS, if the internal energy of a reaction ion is near the critical energies, counterintuitively enormous isotope fractionation will happen, which may explain the remarkably high isotope fractionation from M-$Cl_4$ ion to P-$Cl_3$ of o,p'-DDE observed in this study (Figure 4).

*Effects of Inter-ion and Intra-ion Isotope Fractionations on the Measured Isotope Ratios.* In this study, the measured isotope ratios of molecular ions were only affected by inter-ion isotope fractionation, whereas those of product ions might be influenced by both inter-ion and intra-ion isotope fractionations. Two types of inter-ion fractionations could affect the measured isotope ratio of a certain product ion in opposite directions. The first took place during the dehalogenation from a precursor ion to the product ion, which negatively affected the measured isotope ratio of the product ion. The second occurred during further dehalogenation of the product ion, which positively affected the measured isotope ratio of this product ion. In this situation, the product ion already become a "precursor ion" of its further product ion. Thus, the observed isotope fractionation was apparent rather than real. The direction and extent of the observed apparent isotope fractionation from a precursor ion to its product ion were determined by combinatory effects of these inter-ion and intra-ion isotope fractionations.

The measured isotope ratios of the molecular ions were supposed to be higher than those of total ionized molecular ions due to the inter-ion isotope fractionation. If intra-ion isotope fractionation presented, the measured isotope ratios of the product ions were also deduced to be higher than those of the reacting precursor ions, even though the inter-ion and intra-ion fractionations oppositely contributed to the measured isotope ratios (apparent isotope ratios) of the product ions.[21]

*Proofs for Inter-ion and Intra-ion Isotope Fractionations.* In attempt to reveal relationship between inter-ion and intra-ion isotope fractionations as well as their contribution to the observed isotope fractionation, comprehensive isotope ratio was evaluated with total ions using eq 2 and compared with isotope ratio of molecular ion.

As illustrated in Figure 5, the comprehensive isotope ratios were mostly higher or nearly equal to those of the corresponding molecular ions. Actually, similar phenomenon has been observed in a previous study, yet the authors did not mention any reason.[1] Theoretically, comprehensive isotope ratio of a compound reflects integration of inter-ion and intra-ion isotope fractionations provided intra-ion fractionation occurs in EI-MS. If only inter-ion isotope fractionation occurs in EI-MS, the comprehensive isotope ratio of a compound should be lower than the molecular-ion isotope ratio. For asymmetric compounds, such as PCB-18 and PCB-28, if the differences of reaction rate constants of dehalogenation on different positions are large enough, e.g., at the EI energy of 45 eV, the C-Cl bonds are broken stepwise and thus no intra-ion isotope fractionation presents. In these cases, the molecular-ion isotope ratios are higher than the comprehensive isotope ratios (Figure 5). As the internal energy increases, different C-Cl bonds can be cleaved synchronously, thus intra-ion isotope fractionation can take place. This might be the reason why the molecular-ion isotope ratios of PCB-18 and PCB-28 became lower than the comprehensive isotope ratios at the high EI energy of 70 eV (Figure 5). Another possible reaction would be as follows: once a specific bond was cleaved, the remaining bonds became more symmetric or had smaller differences in critical energies, accordingly leading to synchronous dehalogenation processes. At low internal energies (e.g., with EI energy of 30 eV), the dehalogenation processes became weaker, thus weakening the inter-ion isotope fractionation. In addition, the relative abundances of the molecular ions at low EI energies were higher than those at high EI energies. Therefore, for these asymmetric compounds, the molecular-ion isotope ratios were close to the comprehensive isotope ratios at low internal energies (Figure 5).

Molecular-ion isotope ratio is only impacted by the inter-ion isotope fractionation during the first dehalogenation reaction from the molecular ion to its product ion by losing one Cl/Br atom. Therefore, for a compound at a certain EI energy, if the inter-ion isotope fractionation is dominant, the molecular-ion isotope ratio should be higher than the comprehensive isotope ratio. However, the results indicated that inter-ion isotope fractionation was not dominant in most cases (Figure 5), which means intra-ion isotope fractionation of the HOCs generally occurred in EI-MS and outweighed the inter-ion isotope fractionation. Generation of molecular ions only underwent ionization process by losing an electron, while production of product ions went through dehalogenation reactions involving cleavage of C-Cl/C-Br bonds. Unlike ionization process, dissociation of C-Cl/C-Br bonds could lead to inter-ion and intra-ion isotope fractionations. The measured isotope ratios of product ions were thus possibly impacted by both inter-ion and intra-ion isotope fractionations. In addition, extents of intra-ion isotope fractionation were usually expected to be higher than those of inter-ion isotope fractionation.[13] Accordingly, we deduced that the measured isotope ratios of molecular ions might be more likely to approximate to the true isotope ratios of the HOCs than those of product ions. In addition, due to lack of intra-ion isotope fractionation for molecular ions, the measured isotope ratios of molecular ions presented relatively higher precision than those involving product ions (Table S-3).

**Implication and Prospective Application.** Chlorine and bromine isotope fractionation in EI-MS may negatively affect precision and accuracy of CSIA-Cl/Br results. Therefore, structurally identical isotopic external standards are necessary for CSIA-Cl/Br to compensate the negative effects caused by isotope fractionation in EI-MS. Since EI energy may affect isotope fractionation, stability of EI energy therefore may be crucial to obtain precise CSIA results. In addition, evaluation scheme using complete molecular ions is recommended to perform CSIA-Cl/Br studies to improve the precision and accuracy.

Isotope fractionation may also take place in other conditions, such as in chemical ionization MS, atmospheric-pressure photoionization MS, and collision induced dissociation cell. Therefore, Cl/Br isotope fractionation in analytical instruments should be investigated to ensure accuracy and precision of results. In addition, revealing the Cl/Br isotope fractionation occurring in EI-MS may be useful to predict isotope fractionation of HOCs during other dehalogenation processes and to further explore the dehalogenation pathways. Based on the findings in this study, it can be deduced that the Cl/Br isotope ratios of lower-halogenated HOCs in environment might be influenced by isotope fractionation during dehalogenation processes of higher-halogenated congeners.

## CONCLUSIONS

GC-EI-MS has been proved to be a competent and cost-effective alternative tool for conventional GC-IRMS and GC offline IRMS in CSIA-Cl research. This study systematically investigated the Cl/Br isotope fractionation in EI-MS of 17 HOCs using GC-EI-DFS-HRMS. All the investigated compounds presented isotope fractionation in EI-MS. The isotope fractionation patterns and extents varied by EI energy for most compounds, indicating significant impacts of EI energy on the isotope fractionation, which means stable EI energy may be necessary to achieve precise CSIA-Cl/Br results.

Mechanisms of the isotope fractionation in EI-MS were preliminarily elucidated. For a certain product ion of a halogenated compound, the inter-ion and intra-ion isotope fractionations contributed counteractively to the apparent isotope ratio. The extents of inter-ion isotope fractionation of molecular ions were found to be usually lower than those of intra-ion isotope fractionation of product ions for most HOCs. The scheme using complete molecular-ion isotopologues for isotope ratio evaluation was recommended as the optimal in comparison with others involving product ions, due to that the calculated isotope ratios with this scheme showed highest precision and were deduced to be close to trueness more likely. The QET was applied to interpreting the Cl/Br isotope fractionation of the HOCs in EI-MS.

Implications of isotope fractionation in EI-MS for CSIA studies using GC-EI-MS were proposed according to the findings of this study, for achieving CSIA data with high precision and accuracy. The method and results of this study could be useful to predict Cl/Br isotope fractionation of HOCs in other dehalogenation processes and further explore the dehalogenation pathways.

## ASSOCIATED CONTENT

The *Supporting Information* is available free of charge on the ACS Publications website at http://pending.

## ACKNOWLEDGEMENTS

Some reference standards gifted by Dr. Lianjun Bao and Dr. Man Ren at Guangzhou Institute of Geochemistry are sincerely appreciated. This work was partially supported by the National Natural Science Foundation of China (Grant No. 41603092).

Mass Spectrometry. Unpublished manuscript (Refer to the *Supporting Information for Review Only*).

Page 26

**Figure legends**

**Figure 1.** Schematic diagram of the data processing procedures and isotope ratio calculation scheme using o,p'-DDE as an example. TIC: total ion chromatogram; IR: isotope ratio ($^{37}Cl/^{35}Cl$); P-Cl$_n$: dehalogenated (dechlorinated) product ion possessing $n$ Cl atom(s); M-Cl$_n$: molecular ion with $n$ Cl atoms.

**Figure 2.** Measured isotope ratios of the molecular ions and dehalogenation product ions of all the investigated compounds in EI-MS at the EI energy of 45 eV. Error bars represent the standard derivations (1σ, injection replicates = 5) of the measured isotope ratios ($^{37}Cl/^{35}Cl$ or $^{81}Br/^{79}Br$); P-Cl$_n$: product ion possessing $n$ Cl atom(s); M-Cl$_n$: molecular ion possessing $n$ Cl atoms; P-Br$_n$: product ion possessing $n$ Br atom(s); M-Br$_n$: molecular ion possessing $n$ Br atom(s). G1-G8: compound groups divided based on the isotope ratios and isotope fractionation patterns.

**Figure 3.** Measured chlorine isotope ratios of the dechlorination product ions (Cl$_1$-Cl$_2$) and the dechloromethylation product ions of o,p'-DDD and two DDTs in EI-MS (45 eV). P-Cl$_n$–CM: dechloromethylation product ion possessing $n$ Cl atom(s).

**Figure 4.** Measured isotope ratios of the ions of some representative compounds in EI-MS at different EI energies of 30, 45, 55 and 70 eV.

**Figure 5.** Measured isotope ratios of the molecular ions and total ions of 10 representative compounds (the first 10 panels) in EI-MS at different EI energies (30, 45, 55 and 70 eV) and those of 7 compounds (the last 2 panels) at 45 eV.

**Figures**

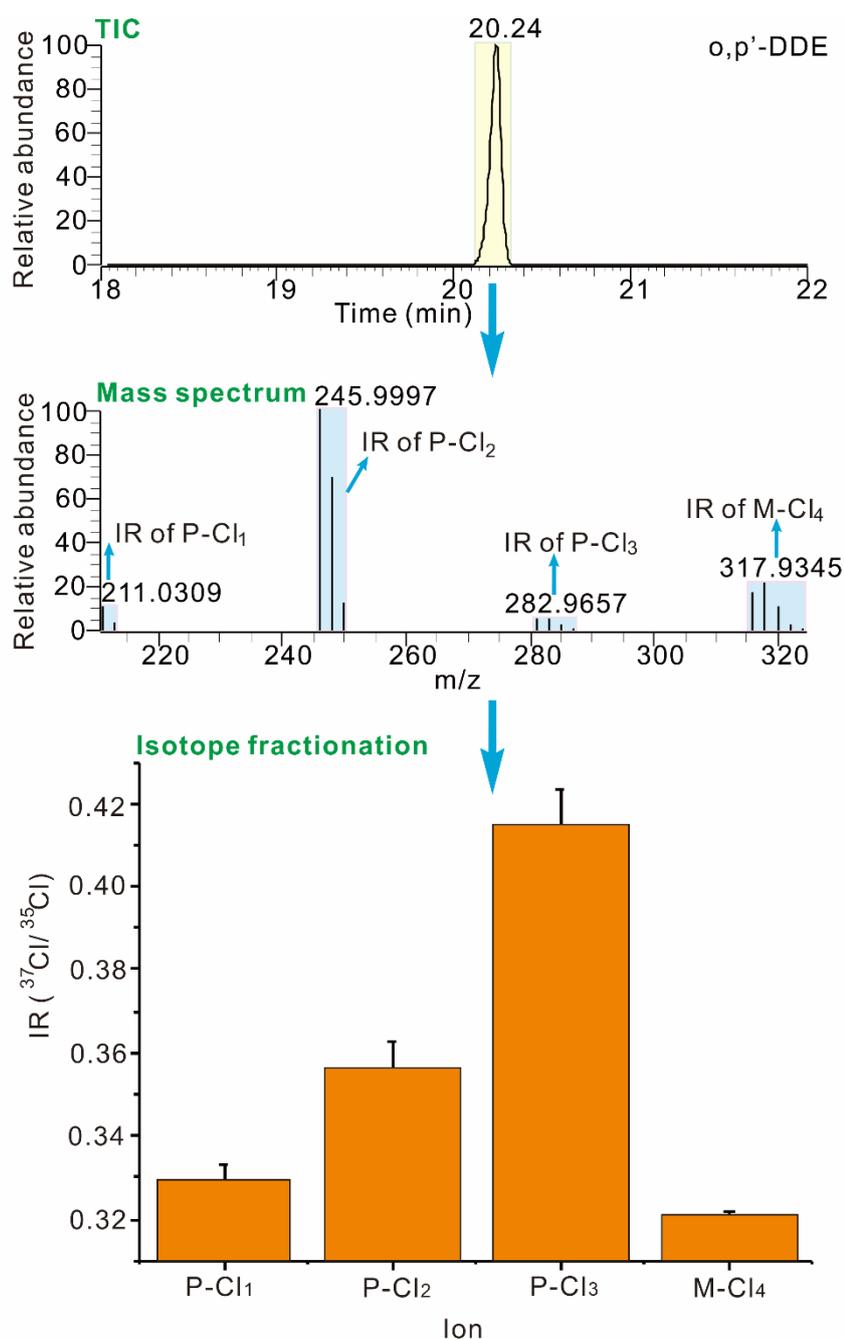

**Figure 1.** Schematic diagram of the data processing procedures and isotope ratio calculation scheme using o,p'-DDE as an example. TIC: total ion chromatogram; IR: isotope ratio ($^{37}Cl/^{35}Cl$); P-$Cl_n$: dehalogenated (dechlorinated) product ion possessing $n$ Cl atom(s); M-$Cl_n$: molecular ion with $n$ Cl atoms; Error bars represent the standard derivations (1σ, injection replicates = 5) of the measured isotope ratios ($^{37}Cl/^{35}Cl$).

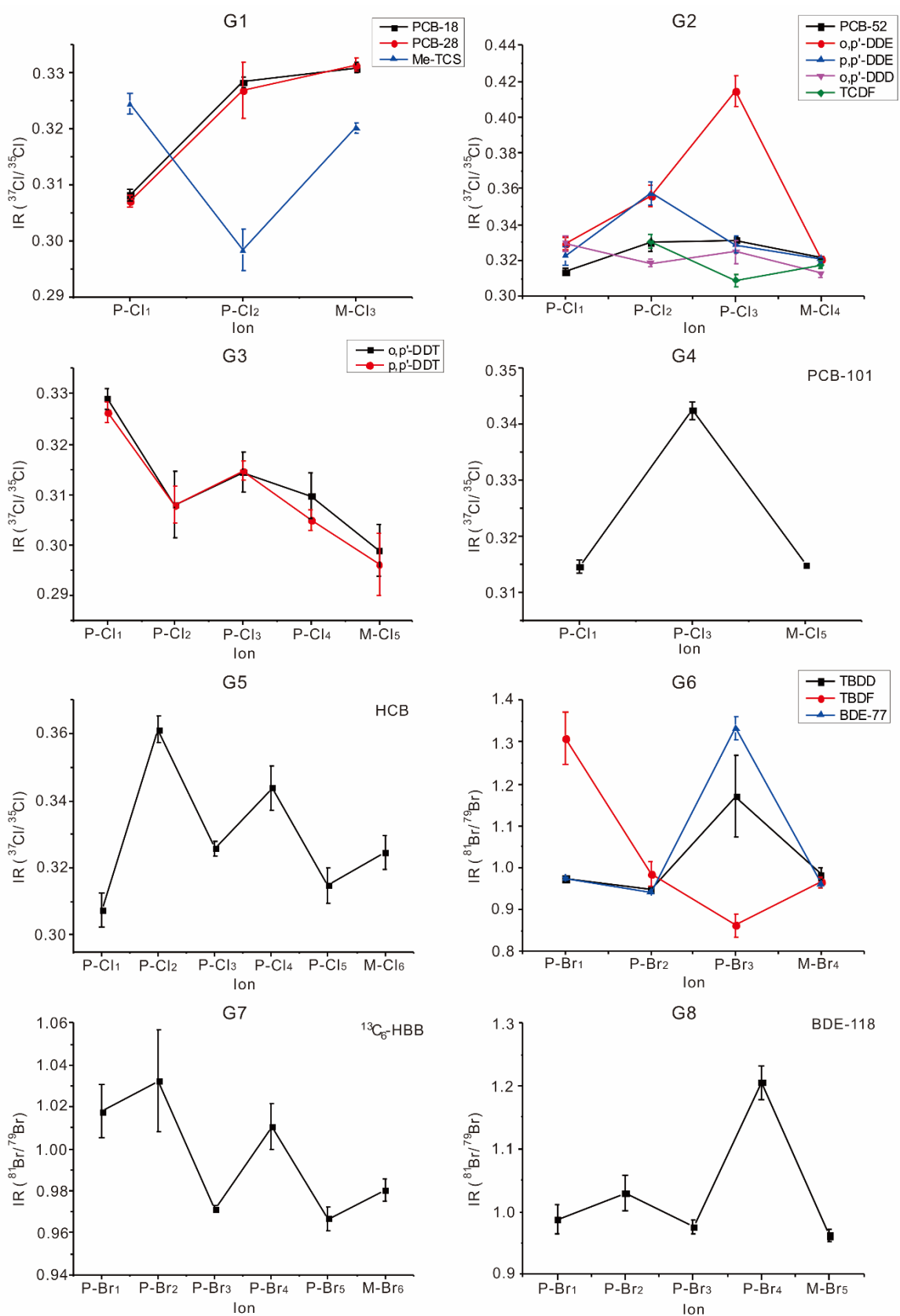

**Figure 2.** Measured isotope ratios of the molecular ions and dehalogenation product ions of all the investigated compounds in EI-MS at the EI energy of 45 eV. Error bars represent the standard derivations (1σ, injection replicates = 5) of the measured isotope ratios ($^{37}Cl/^{35}Cl$ or $^{81}Br/^{79}Br$); IR: isotope ratio ($^{37}Cl/^{35}Cl$ or $^{81}Br/^{79}Br$); P-Cl$_n$: product ion possessing $n$ Cl atom(s);

M-Cl$_n$: molecular ion possessing $n$ Cl atoms; P-Br$_n$: product ion possessing $n$ Br atom(s); M-Br$_n$: molecular ion possessing $n$ Br atom(s). G1-G8: compound groups divided based on the isotope ratios and isotope fractionation patterns.

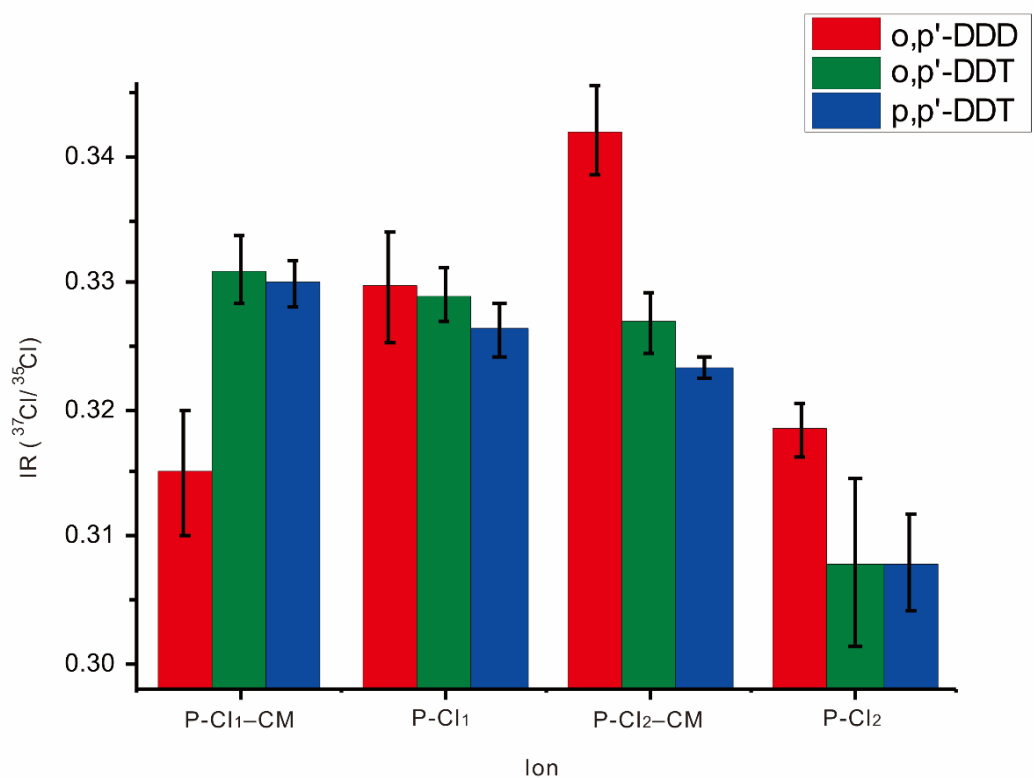

**Figure 3.** Measured chlorine isotope ratios of the dechlorination product ions (Cl$_1$-Cl$_2$) and the dechloromethylation product ions of o,p'-DDD and two DDTs in EI-MS (45 eV). P-Cl$_n$–CM: dechloromethylation product ion possessing $n$ Cl atom(s).

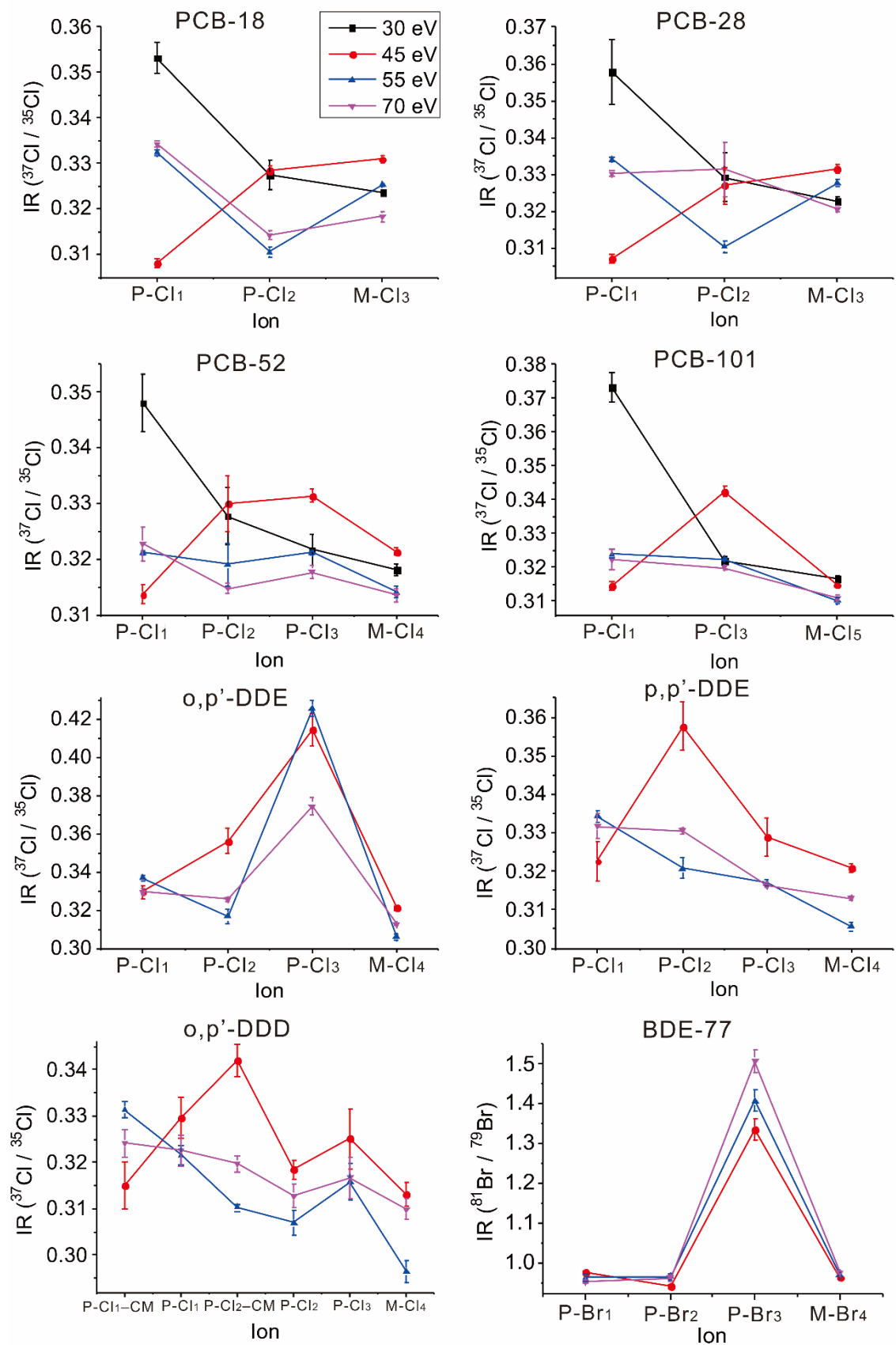

**Figure 4.** Measured isotope ratios of the ions of some representative compounds in EI-MS at different EI energies of 30, 45, 55 and 70 eV.

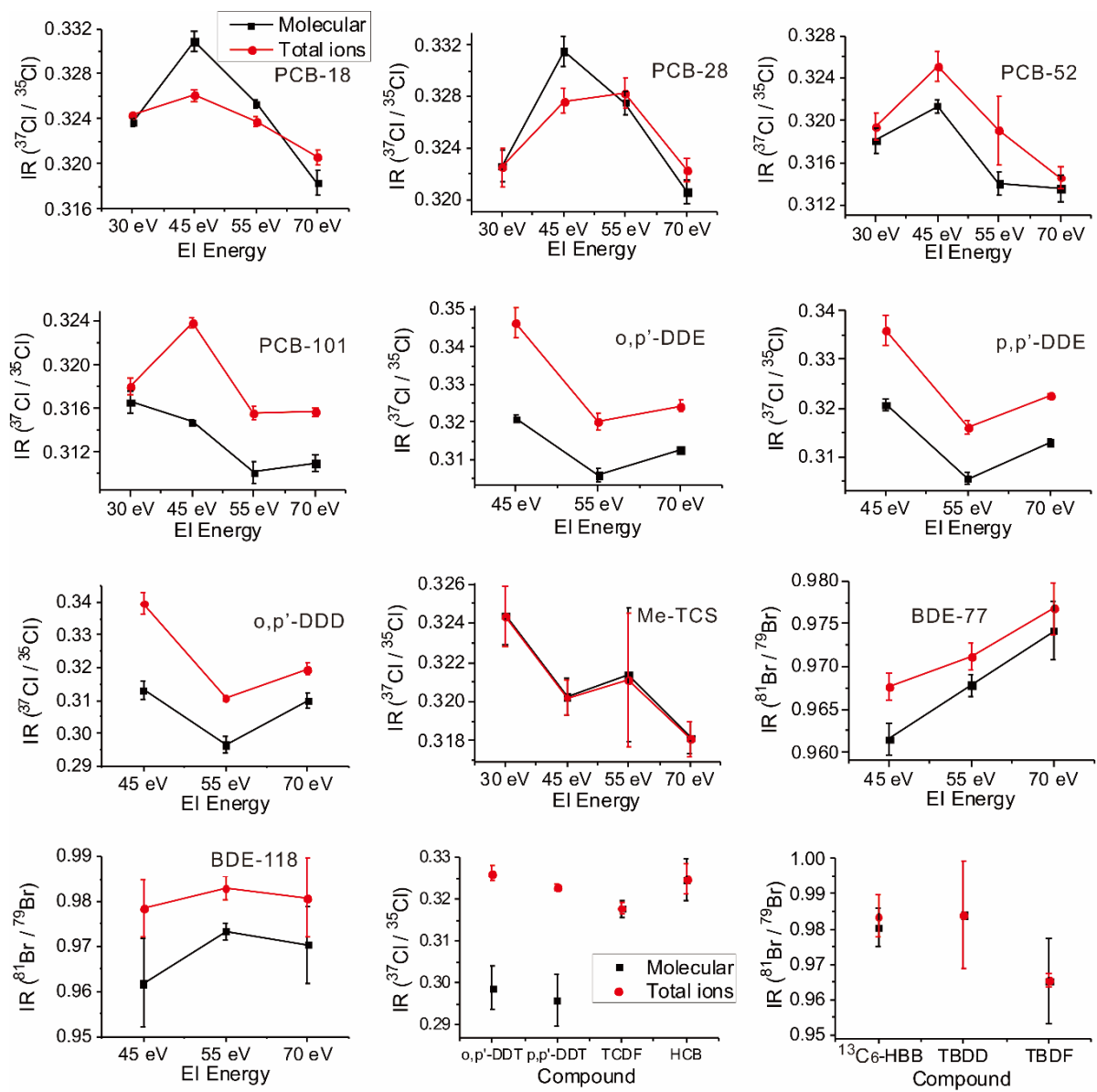

**Figure 5.** Measured isotope ratios of the molecular ions and total ions of 10 representative compounds (the first 10 panels) in EI-MS at different EI energies (30, 45, 55 and 70 eV) and those of 7 compounds (the last 2 panels) at 45 eV.

For TOC only

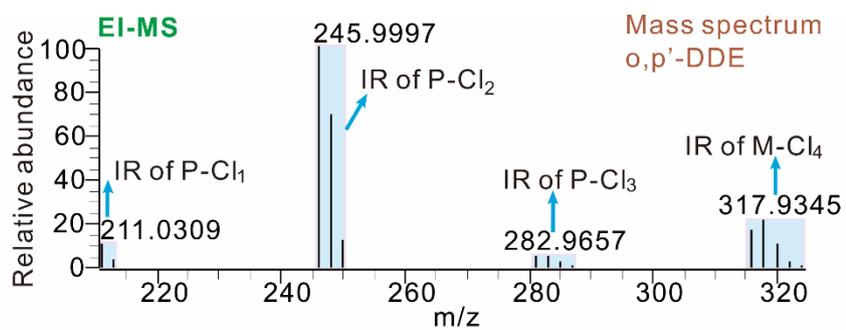
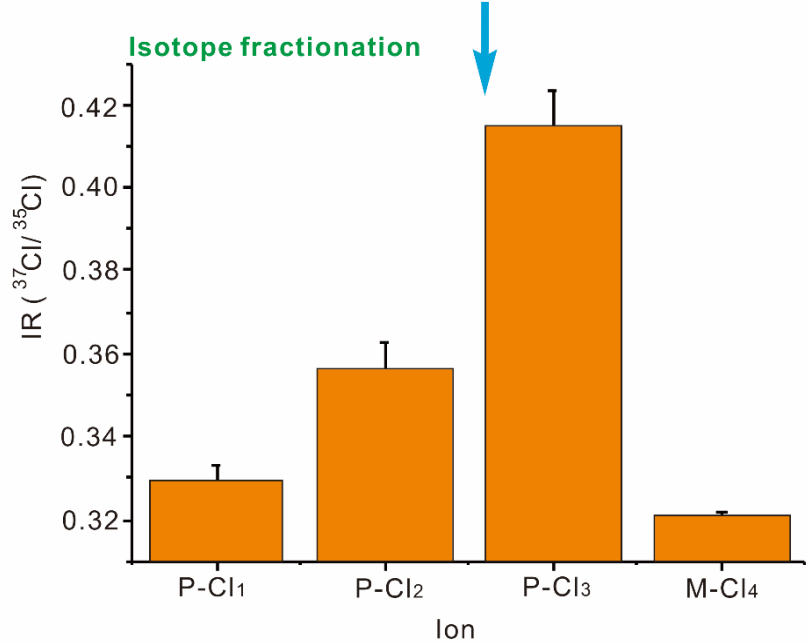